\documentclass[11pt]{article}
\input epsf
\usepackage{psfrag}
\usepackage{graphicx}

\catcode`\@=11
\def\@citex[#1]#2{%
\if@filesw \immediate \write \@auxout {\string \citation {#2}}\fi
\@tempcntb\m@ne \let\@h@ld\relax \def\@citea{}%
\@cite{%
  \@for \@citeb:=#2\do {%
    \@ifundefined {b@\@citeb}%
      {\@h@ld\@citea\@tempcntb\m@ne{\bf ?}%
      \@warning {Citation `\@citeb ' on page \thepage \space undefined}}%
      {\@tempcnta\@tempcntb \advance\@tempcnta\@ne%
      \@tempcntb\number\csname b@\@citeb \endcsname \relax%
      \ifnum\@tempcnta=\@tempcntb 
        \ifx\@h@ld\relax%
          \edef \@h@ld{\@citea\csname b@\@citeb\endcsname}%
        \else%
          \edef\@h@ld{\ifmmode{-}\else--\fi\csname b@\@citeb\endcsname}%
        \fi%
      \else
        \@h@ld\@citea\csname b@\@citeb \endcsname%
        \let\@h@ld\relax%
      \fi}%
    \def\@citea{,\penalty\@highpenalty\,}%
  }\@h@ld
}{#1}}

\def\@citeb#1#2{{[#1]\if@tempswa , #2\fi}}
\def\@citeu#1#2{{$^{#1}$\if@tempswa , #2\fi }}
\def\@citep#1#2{{#1\if@tempswa , #2\fi}}

\def\bcites{         
        \catcode`\@=11
        \let\@cite=\@citeb
        \catcode`\@=12
}

\def\upcites{         
        \catcode`\@=11
        \let\@cite=\@citeu
        \catcode`\@=12
}

\def\plaincites{      
        \catcode`\@=11
        \let\@cite=\@citep
        \catcode`\@=12
}

\newcount\hour
\newcount\minute
\newtoks\amorpm
\hour=\time\divide\hour by 60
\minute=\time{\multiply\hour by 60 \global\advance\minute by-\hour}
\edef\standardtime{{\ifnum\hour<12 \global\amorpm={am}%
        \else\global\amorpm={pm}\advance\hour by-12 \fi
        \ifnum\hour=0 \hour=12 \fi
        \number\hour:\ifnum\minute<10 0\fi\number\minute\the\amorpm}}
\edef\militarytime{\number\hour:\ifnum\minute<10 0\fi\number\minute}

\def\draftlabel#1{{\@bsphack\if@filesw {\let\thepage\relax
   \xdef\@gtempa{\write\@auxout{\string
      \newlabel{#1}{{\@currentlabel}{\thepage}}}}}\@gtempa
   \if@nobreak \ifvmode\nobreak\fi\fi\fi\@esphack}
        \gdef\@eqnlabel{#1}}
\def\@eqnlabel{}
\def\@vacuum{}
\def\marginnote#1{}
\def\draftmarginnote#1{\marginpar{\raggedright\scriptsize\tt#1}}
\def\draft{
        \pagestyle{plain}
        \overfullrule=2pt
        \oddsidemargin -.5truein
        \def\@oddhead{\sl \phantom{\today\quad\militarytime} \hfil
        \smash{\Large\sl DRAFT} \hfil \today\quad\militarytime}
        \let\@evenhead\@oddhead
        \let\label=\draftlabel
        \let\marginnote=\draftmarginnote
        \def\ps@empty{\let\@mkboth\@gobbletwo
        \def\@oddfoot{\hfil \smash{\Large\sl DRAFT} \hfil}
        \let\@evenfoot\@oddhead}
        \def\@eqnnum{(\theequation)\rlap{\kern\marginparsep\tt\@eqnlabel}%
        \global\let\@eqnlabel\@vacuum}  }

\def\blackfonts{
        \font\blackboard=msbm10 scaled\magstep1
        \font\blackboards=msbm8
        \font\blackboardss=msbm6
}

\def\prep{         
        \catcode`\@=11
        \input art10.sty
        \catcode`\@=12
        
        \let\small\null
        \def\blackfonts{
                \font\blackboard=msbm10
                \font\blackboards=msbm7
                \font\blackboardss=msbm5
        }
        \let\sl\it
        \twocolumn
        \sloppy
        \voffset=-2.54truecm
        \hoffset=-2.54truecm
        \flushbottom
        \parindent 1em
        \leftmargini 2em
        \leftmarginv .5em
        \leftmarginvi .5em
        \marginparwidth 48pt
        \marginparsep 10pt
        \setlength{\columnsep}{2truecm}
        \setlength{\textwidth}{25.4truecm}
        \setlength{\textheight}{17truecm}
        \baselineskip=16pt
        \oddsidemargin .18truein
        \evensidemargin .17truein
}

\def\eqalign#1{\null\,\vcenter{\openup\jot\m@th
  \ialign{\strut\hfil$\displaystyle{##}$&$\displaystyle{{}##}$\hfil
      \crcr#1\crcr}}\,}
\def\eqalignno#1{\displ@y \tabskip\centering
  \halign to\displaywidth{\hfil$\@lign\displaystyle{##}$\tabskip\z@skip
    &$\@lign\displaystyle{{}##}$\hfil\tabskip\centering
    &\llap{$\@lign##$}\tabskip\z@skip\crcr
    #1\crcr}}

\def\section{\@startsection {section}{1}{\z@}{3.ex plus 1ex minus
 .2ex}{2.ex plus .2ex}{\large\bf}}
\def\subsection{\@startsection{subsection}{2}{\z@}{2.75ex plus 1ex minus
 .2ex}{1.5ex plus .2ex}{\bf}}        

\def\appendix{{\newpage\section*{Appendix}}\let\appendix\section%
        {\setcounter{section}{0}
        \gdef\thesection{\Alph{section}}}\section}

\def\abstract{\if@twocolumn
\section*{Abstract}
\else 
\begin{center}
{\bf Abstract\vspace{-.5em}\vspace{0pt}}
\end{center}
\quotation
\fi}

\catcode`\@=12

\def\sqr#1#2{{\vcenter{\vbox{\hrule height.#2pt\hbox{\vrule width.#2pt 
height#1pt \kern#1pt \vrule width.#2pt}\hrule height.#2pt}}}}

\def\=d{\,{\buildrel\rm def\over =}\,}

\def\i3p{\p32\int d^3p}

\def\As{A\hbox to 1pt{\hss /}}
\def\np4{\int d^4p_1\cdots d^4p_{n-1}\, }

\def\nx4{\int d^4x_1\ldots d^4x_n\, }

\def\kon#1#2{\vbox{\halign{##&&##\cr
\lower4pt\hbox{$\scriptscriptstyle\vert$}\hrulefill &
\hrulefill\lower4pt\hbox{$\scriptscriptstyle\vert$}\cr $#1$&
$#2$\cr}}}

\def\konv#1#2#3{\hbox{\vrule height12pt depth-1pt}
\vbox{\hrule height12pt width#1cm depth-11.6pt}
\hbox{\vrule height6.5pt depth-0.5pt}
\vbox{\hrule height11pt width#2cm depth-10.6pt\kern5pt
      \hrule height6.5pt width#2cm depth-6.1pt}
\hbox{\vrule height12pt depth-1pt}
\vbox{\hrule height6.5pt width#3cm depth-6.1pt}
\hbox{\vrule height6.5pt depth-0.5pt}}
\def\konu#1#2#3{\hbox{\vrule height12pt depth-1pt}
\vbox{\hrule height1pt width#1cm depth-0.6pt}
\hbox{\vrule height12pt depth-6.5pt}
\vbox{\hrule height6pt width#2cm depth-5.6pt\kern5pt
      \hrule height1pt width#2cm depth-0.6pt}
\hbox{\vrule height12pt depth-6.5pt}
\vbox{\hrule height1pt width#3cm depth-0.6pt}
\hbox{\vrule height12pt depth-1pt}}

\def\konw#1#2#3{\hbox{\vrule height12pt depth-1pt}
\vbox{\hrule height12pt width#1cm depth-11.6pt}
\hbox{\vrule height6.5pt depth-0.5pt}
\vbox{\hrule height12pt width#2cm depth-11.6pt \kern5pt
      \hrule height6.5pt width#2cm depth-6.1pt}
\hbox{\vrule height6.5pt depth-0.5pt}
\vbox{\hrule height12pt width#3cm depth-11.6pt}
\hbox{\vrule height12pt depth-1pt}}

\def\i{{\rm int}}

\def\e{{\rm ext}}

\def\a{{\rm av}}

\def\m3{{\mu_1\mu_2\mu_3}}

\def\p{{(+)}}

\def\be{\begin{equation}}       \def\eq{\begin{equation}}
\def\ee{\end{equation}}         \def\eqe{\end{equation}}

\def\bea{\begin{eqnarray}}      \def\eqa{\begin{eqnarray}}
\def\ena{\end{eqnarray}}        \def\eea{\end{eqnarray}}
                                \def\eqae{\end{eqnarray}}

\def\ba{\begin{array}}
\def\ea{\end{array}}
\def\unit{1 \hskip-.3em \raise2pt\hbox{$ \scriptstyle |$ } }

\def\a{\alpha}

\def\e{\epsilon}           

\def\i{\iota}

\def\m{\mu}

\def\p{\pi}                
\def\t{\tau}

\def\D{\Delta}

\def\bop#1{\setbox0=\hbox{$#1M$}\mkern1.5mu
        \vbox{\hrule height0pt depth.04\ht0
        \hbox{\vrule width.04\ht0 height.9\ht0 \kern.9\ht0
        \vrule width.04\ht0}\hrule height.04\ht0}\mkern1.5mu}

\def\>{\rangle} 

\def\<{\langle} 
\def\Dsl{D \hskip-.6em \raise1pt\hbox{$ / $ } }

\def\sl#1{\rlap{\hbox{$\mskip 1 mu /$}}#1}
\def\leftrightarrowfill{$\mathsurround=0pt \mathord\leftarrow \mkern-6mu
       \cleaders\hbox{$\mkern-2mu \mathord- \mkern-2mu$}\hfill
       \mkern-6mu \mathord\rightarrow$}
\def\dvec#1{\vbox{\ialign{##\crcr
       \leftrightarrowfill\crcr\noalign{\kern-1pt\nointerlineskip}
       $\hfil\displaystyle{#1}\hfil$\crcr}}}          
\def\hook#1{{\vrule height#1pt width0.4pt depth0pt}}
\def\leftrighthookfill#1{$\mathsurround=0pt \mathord\hook#1
       \hrulefill\mathord\hook#1$}
\def\underhook#1{\vtop{\ialign{##\crcr                 
       $\hfil\displaystyle{#1}\hfil$\crcr
       \noalign{\kern-1pt\nointerlineskip\vskip2pt}
       \leftrighthookfill5\crcr}}}
\def\smallunderhook#1{\vtop{\ialign{##\crcr      
       $\hfil\scriptstyle{#1}\hfil$\crcr
       \noalign{\kern-1pt\nointerlineskip\vskip2pt}
       \leftrighthookfill3\crcr}}}

\def\sfrac#1#2{{\vphantom1\smash{\lower.5ex\hbox{\small$#1$}}\over
       \vphantom1\smash{\raise.4ex\hbox{\small$#2$}}}} 
\def\bfrac#1#2{{\vphantom1\smash{\lower.5ex\hbox{$#1$}}\over
       \vphantom1\smash{\raise.3ex\hbox{$#2$}}}}      
\def\afrac#1#2{{\vphantom1\smash{\lower.5ex\hbox{$#1$}}\over#2}}  
\def\on#1#2{{\buildrel{\mkern2.5mu#1\mkern-2.5mu}\over{#2}}}
\def\ddt#1{\on{\hbox{\LARGE .\kern-2pt.}}#1}             
\def\tdt#1{\on{\hbox{\LARGE .\kern-2pt.\kern-2pt.}}#1}   

\def\boxes#1{
       \newcount\num
       \num=1
       \newdimen\downsy
       \downsy=-1.5ex
       \mskip-2.8mu
       \bo
       \loop
       \ifnum\num<#1
       \llap{\raise\num\downsy\hbox{$\bo$}}
       \advance\num by1
       \repeat}
\def\boxup#1#2{\newcount\numup
       \numup=#1
       \advance\numup by-1
       \newdimen\upsy
       \upsy=.75ex
       \mskip2.8mu
       \raise\numup\upsy\hbox{$#2$}}

\newskip\humongous \humongous=0pt plus 1000pt minus 1000pt
\def\caja{\mathsurround=0pt}
\def\eqalign#1{\,\vcenter{\openup2\jot \caja
       \ialign{\strut \hfil$\displaystyle{##}$&$
       \displaystyle{{}##}$\hfil\crcr#1\crcr}}\,}
\newif\ifdtup

\def\1ov4{{1\over 4}}

\def\ddt{\dot{\t}}

\renewcommand{\a}{\alpha}

\newcommand{\rmd}{{\rm d}}

\newcommand{\beq}{\begin{equation}}
\newcommand{\eeq}{\end{equation}}
\newcommand{\tD}{{\tilde{\Delta}}}
\def\ba{\begin{eqnarray}}
\def\ea{\end{eqnarray}}

\begin{document}



\null\vskip-24pt
\hfill KL-TH 00/01
\vskip-10pt
\hfill {\tt hep-th/0002025}
\vskip0.3truecm
\begin{center}
\vskip 3truecm
{\Large\bf
A note on the analyticity of AdS scalar exchange graphs in the crossed channel
}\\ 
\vskip 1.5truecm
{\large\bf
Laurent Hoffmann, 
 \footnote{email:{\tt hoffmann@physik.uni-kl.de}} Anastasios C. Petkou
   \footnote{email:{\tt
       petkou@physik.uni-kl.de}}   
and Werner R\" uhl \footnote{
email:{\tt ruehl@physik.uni-kl.de}}
}\\
\vskip 1truecm
{\it Department of Physics, Theoretical Physics\\
University of Kaiserslautern, Postfach 3049 \\
67653 Kaiserslautern, Germany}\\

\end{center}
\vskip 3truecm
\centerline{\bf Abstract}

We discuss the analytic properties of AdS scalar exchange
graphs in the crossed channel. We show that the possible non-analytic 
terms drop out by virtue of 
non-trivial properties of generalized hypergeometric functions. The
absence of non-analytic terms is
a necessary condition for the existence of an operator product
expansion for CFT amplitudes obtained from AdS/CFT correspondence.

\newpage

Scalar exchange conformal graphs are basic ingredients in a
calculation of conformally invariant $n$-point functions for $n\geq 
4$. Such graphs have been extensively studied in earlier works on  
conformal field theory (CFT) in $d>2$ \cite{fradkin,ruehl1,tassos1}.
Consider a conformal four-point function \footnote{For clarity we
  choose  all fields in the four-point function to have equal
  dimensions. The generalization of our results in the case of 
  different dimensions is straightforward.} ${\cal G}(x_1,..,x_4)= \langle{\cal
  O}(x_1)..{\cal O}(x_4)\rangle$ where the scalar field ${\cal O}(x)$
has dimension $\tilde{\Delta}$. One contribution to this
four-point function is given by the skeleton graphs in which a scalar
field $\sigma(x)$  of dimension $\D$ is exchanged between
external legs corresponding to ${\cal O}(x)$. For such skeleton graphs
to exist  it suffices that the three-point
function $\langle {\cal O}(x_1){\cal O}(x_2)\sigma(x_3)\rangle $ is
non-zero. \footnote{Conformal invariance determines the form of
three-point functions. Then, using the D'EPP formula
\cite{depp} for amputation one can obtain the full three-point
vertex function which is used in the skeleton graphs.} 
The relevant graph is depicted in Fig.1. 
Then, bose symmetry requires that ${\cal G}(x_1,..,x_4)$ 
receives contributions from all different graphs which can be obtained
from Fig.1  by suitable relabeling  of the external points.

\begin{figure}[h]
\centering
\begin{minipage}{14cm}
\centering
\psfrag{x1}{$x_1$}
\psfrag{x2}{$x_2$}
\psfrag{x3}{$x_3$}
\psfrag{x4}{$x_4$}
\psfrag{C}{$B(x_1,x_3;x_2,x_4)\,=$}
\includegraphics[width=7cm]{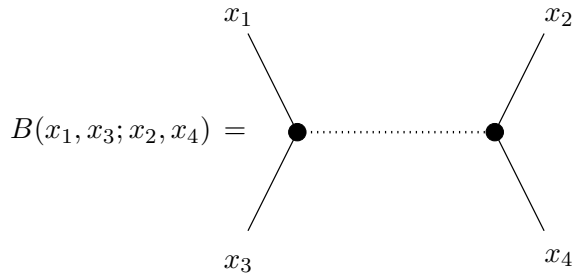}
\centering
\caption{{\it \small The standard CFT$_{d}$ scalar exchange graph. The
    solid lines 
  represent the full propagator of ${\cal O}(x)$, the dotted line
  represents the full propagator of $\sigma(x)$ and the dark blobs
  stand for the full vertex functions }\cite{ruehl1,tassos1}.}
\end{minipage}
\end{figure} 

One of the basic implications of conformal invariance is the existence
of an operator product expansion (OPE) \cite{mack}  which is
essentially equivalent to a partial-wave decomposition of conformal
$n$-point functions. Such a property 
finds an  explicit application in the study of the four-point
function ${\cal 
  G}(x_1,..,x_4)$, as it implies that its calculation in terms of
skeleton graphs should be compatible with a conformal OPE.  By
conformal invariance ${\cal G}(x_1,..,x_4)$  is determined up to 
an arbitrary analytic function of the  two biharmonic ratios
\beq
u=\frac{x_{13}^2x_{24}^2}{x_{12}^2 x_{34}^2}\,\,\,\,,\,\,\,\,
v=\frac{x_{14}^2x_{23}^2}{x_{12}^2 x_{34}^2}\,,\,\,\,\, x_{ij}=x_i-x_j
\,.
\eeq
Then, calculating the relevant skeleton graphs obtained
from  Fig.1 one finds a general expansion of the form 
\ba
\label{gen4pt}
{\cal G}^{(1)}(x_1,x_2,x_3,x_4) &=&
\frac{1}{(x_{12}^2x_{34}^2)^{\tilde{\Delta}}}
\sum_{n,m=0}^{\infty}\frac{u^n
  (1-v)^m}{n!m!}\Biggl[ -A_{nm}(\tD,\D) \ln u +B_{nm}(\tD,\D) \nonumber \\
 & & \hspace{2cm}+u^{\frac{1}{2}\Delta
 -\tilde{\Delta}} C_{nm}(\tilde{\Delta},\Delta) 
+u^{\frac{1}{2}(d-\Delta)-\tilde{\Delta}}  
C_{nm}(\tilde{\Delta},d-\Delta) \Biggl] \,,
\ea
where the coefficients $A_{nm}$, $B_{nm}$ and $C_{nm}$ have
been given in a number of works \cite{ruehl1,tassos1}. This expression
is compatible with a conformal OPE when the latter is 
inserted in the {\it direct channel} $x_1\rightarrow x_3$,
$x_2\rightarrow x_4$ or equivalently $u\rightarrow 0$, $v\rightarrow
1$. The first two terms on the r.h.s. of (\ref{gen4pt}) together with
the contributions from the zeroth order terms (disconnected graphs)
\footnote{Here and in the following we set to 1 the normalization of
  two-point functions.}
\beq
\label{disc}
{\cal G}^{(0)}(x_1,x_2,x_3,x_4) =
\frac{1}{(x_{13}^2x_{24}^2)^{\tilde{\Delta}}} +
\frac{1}{(x_{12}^2x_{34}^2)^{\tilde{\Delta}}}
\left[1+v^{-\tilde{\Delta}}\right]\,, 
\eeq
represent 
the contributions of infinitely many conformal tensor fields with
definite dimension and rank. The last two terms on the r.h.s. of
(\ref{disc}) determine the canonical dimensions of these tensor
fields, while the logarithmic terms on the
r.h.s. of (\ref{gen4pt}) determine their anomalous dimension. A general
method to identify and study 
the contributions of conformal tensor fields in four-point functions is
described in \cite{HPR}. 

In order that a four-point function admits a conformal OPE
it is necessary that it does not contain non-analytic
terms e.g. the double summation in 
(\ref{gen4pt}) is restricted over
$n,m\in{\bf N}_0$. The 
analyticity of conformally
invariant amplitudes is a non-trivial property of CFT$_d$ and should
be checked in all explicit calculations. In this letter
we show that general AdS$_{d+1}$ scalar
exchange graphs are analytic in the crossed channel expansion and
therefore 
admit a conformal OPE. More technical details of
our calculation as well as a thorough OPE analysis of AdS$_{d+1}$
scalar exchange graphs are
contained in  \cite{HPR}. 

Our main motivation for studying the analyticity properties of AdS$_{d+1}$
scalar exchange graphs comes from similar studies of 
conformal graphs such as the one depicted in Fig.1. Using the D'EPP
formula and Symanzik's technique \cite{symanzik} the exchange graph
in Fig.1 has been calculated in the direct channel as
\ba
\label{dchann}
B(x_1,x_3;x_2,x_4) & =
&\frac{\pi^du^{-\tD}}{(x_{12}^2x_{34}^2)^{\tD}}\sum_{n,m=0}^{\infty}
\frac{u^n(1-v)^m}{n!m!}\Bigl[u^{\frac{1}{2}\D}c_{nm}(d,\D)
+u^{\frac{1}{2}d-\frac 
  {1}{2}\D} c_{nm}(d,d-\D)\Bigl]\!,  \\
c_{nm}(d,\D)&= & \a(\D)\a^2(\sfrac12 d- \sfrac12\D)
\frac{(\frac{1}{2}\D)_n^2
  (\frac{1}{2}\D)_{n+m}^2}{(\D)_{2n+m}(\D-\frac{1}{2}d+1)_n}\,,\\
\a(b) & = &\frac{\Gamma(\frac12 d-b)}{\Gamma(b)} \,,\,\,\,\, (b)_n
\,\, = \,\, \frac{\Gamma(b+n)}{\Gamma(b)}\,.
\ea
This result is explicitly compatible with a
conformally invariant OPE when the latter is inserted in the direct
channel. Namely, the first term on the r.h.s. of  (\ref{dchann}) is
the full contribution 
of the scalar field $\sigma(x)$ in the four-point function, while the second
term on the r.h.s. represents the so-called {\it shadow symmetric}
  singularities of the first series. \footnote{The term {\it shadow
    symmetry} was 
  introduced for the first time in \cite{parisi}. It corresponds to
  an intertwiner \cite{koller} of the conformal group in $d>2$ that
  maps the equivalent representations with dimensions $\eta$ and
  $d-\eta$ onto each other. Shadow symmetric singularities
  may correspond to physical {\it shadow fields} if the dimensions of
  the latter satisfy
  the unitarity bound e.g. $d-\eta\geq d/2-1$. See 
  \cite{ruehl1,tassos1} and also \cite{klebanov}.} From (\ref{dchann})
we can obtain the 
results for both the  {\it crossed channels}. For clarity we consider
here one of two crossed channels, namely when we set
$x_3\leftrightarrow x_4$ in Fig.1. This is obtained from
(\ref{dchann}) by the interchange  $u\leftrightarrow v$. However,
since we are interested in comparing the result with the OPE
inserted in the direct channel, the above crossing transformation
requires an analytic continuation of the result
in (\ref{dchann}). Indeed, the hypergeometric $m$-sum on the r.h.s. of
(\ref{dchann}) gives, by virtue of the crossing transformation, a
hypergeometric series in the variable $(1-u)$. In order to compare this
with the OPE inserted in the direct channel we use a degenerate Kummer
transformation \cite{grad} to obtain a series in the variable
$u$. Nevertheless, the remaining series is now a series in the
variable $v$ and it is not suitable for a comparison with the direct
channel OPE which requires $v\rightarrow 1$. Eventually, we need to
transform this series into a power series in $(1-v)$ and this step
requires an analytic continuation which may not always be
possible. For the explicit case of 
(\ref{dchann}) 
we obtain after some algebra,
\ba
\label{cchann}
B(x_1,x_4;x_2,x_3) & = &K\frac{1}{(x_{12}^2x_{34}^2)^{\tD}}
\sum_{n,m=0}^{\infty} \frac{u^m}{(m!)^2}{\cal
  D}_m(\frac{\partial}{\partial\xi})\Bigl[(1-v)^{s-2t} {\cal A}_{m}(\xi) +
{\cal B}_{m}(\xi)\Bigl]_{\xi =0}\,,\\
{\cal A}_{m}(\xi) & = & v^{\frac{1}{2}\D}\Gamma(2t-s)
\,{}_2F_1\Bigl(s-t,s-t; 1+s-2t;1-v \Bigl)
  \nonumber \\
 & & -v^{\frac12 d-\frac12 \D}\Gamma(2t-s)
\,{}_2F_1\Bigl(1-t,1-t;1+s-2t;1-v) \label{nan}\,,\\
{\cal B}_m(\xi) & = & v^{\frac12\D}\,\Gamma(s-2t)
\,_2F_1\Bigl(t,t;2t-s+1;1-v\Bigl)\left[\frac{\Gamma^2( t)}{\Gamma^2(s-t)}
  -\frac{\Gamma^2( 1+t-s)}{\Gamma^2(1-t)}\right]\label{an}\!\!,
\ea
where 
\ba
s & = & \D-\sfrac12 d+1 \,,\, t\,= \,\sfrac12 \D +m+\xi\,,\\
{\cal D}_m(\frac{\partial}{\partial\xi}) & = & -\ln u +2\psi(m+1)
  -\sfrac{\partial}{\partial \xi}\,,  
\ea
and $K$ is an overall constant whose explicit value is not important
here \cite{HPR}.
Notice that the second term on the r.h.s. of (\ref {cchann}) is
analytic, while the first is not. Nevertheless, the non-analytic
term is exactly zero as can be easily seen using Euler's identity
\cite{grad} 
for either of the two  hypergeometric functions in (\ref{nan}). The
remaining analytic term can then be easily computed from (\ref{cchann})
and (\ref{an}) and the result was given in \cite{ruehl1,tassos1}. 

\begin{figure}[t]
\centering
\begin{minipage}{14cm}
\centering
\psfrag{x1}{$x_1$}
\psfrag{x2}{$x_2$}
\psfrag{x3}{$x_3$}
\psfrag{x4}{$x_4$}
\psfrag{C}{$C(x_1,x_3;x_2,x_4)=$}
\includegraphics[width=7cm]{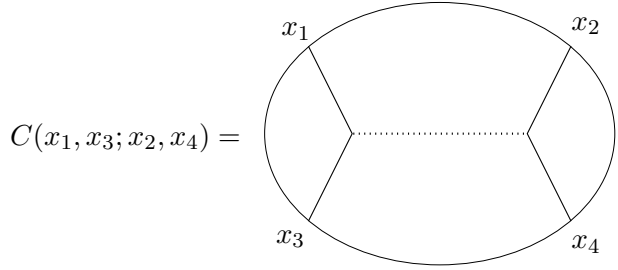}
\centering
\caption{{\it \small The AdS$_{d+1}$ scalar exchange graph. The solid
    lines represent  
  the ``bulk-to-boundary'' propagator of ${\cal O}(x)$ and the dotted
  line represents the standard ``bulk-to-bulk'' propagator of
  $\sigma(x)$ }\cite{freedman1}.}  
\end{minipage}
\end{figure}

In the above calculation we observe that the existence of the shadow
singularities in the direct 
channel 
(\ref{dchann}) was essential for the cancellation of the non-analytic
terms in the crossed channel.  Let us turn now to the calculation of
the standard AdS$_{d+1}$ scalar exchange graph \cite{freedman1,liu}
depicted in Fig.2. 
This, 
together with the two crossed symmetric graphs  and the
three disconnected terms (\ref{disc}) contribute to the four-point
function of ${\cal O}(x)$ in the boundary CFT$_d$ in the context of AdS/CFT
correspondence \cite{maldacena}. Our starting point is the following
Mellin-Barnes representation for the graph in Fig.2 which is obtained
using either Symanzik's method \cite{HPR} or by other techniques \cite{liu}
\ba
\label{mb}
C(x_1,x_3;x_2,x_4) & = & \frac{k}{(x_{12}^2x_{34}^2)^{\tD}}
\int_{{\cal C}}\frac{\rmd s}{2\pi{\rm i}} \Gamma^2(-s) \Biggl[
\frac{\Gamma^4(\tD+s) \Gamma^2(\frac12 \D+\tD-\frac12 d)
  \Gamma(\frac12 \D-\tD-s)}{ \Gamma(2\tD +2s)\Gamma(\D-\frac12
  d+1)\Gamma(\frac12 \D+\tD-\frac12 d-s)} \nonumber \\
&& \hspace{-2.5cm} \times _3F_2\left( \sfrac12 \D+\tD-\sfrac12
  d,\sfrac12 \D+\tD-\sfrac12 
  d,\sfrac12 \D-\tD-s; \D-\sfrac12 d +1,\tD+\sfrac12\D-\sfrac12
  d-s;1\right) \nonumber \\
& & \hspace{3cm}\times u^s \,_2F_1\left(\tD +s,\tD +s;2\tD +2s;
  1-v\right)\Biggl]\,, 
\ea  
where $k$ is an unimportant for what follows constant
\cite{HPR}.  The contour ${\cal C}$ is chosen such that it separates
the ``right'' 
from the ``left'' poles of the integrand \cite{liu,titsch}.  The existence of
such a contour is 
necessary for the convergence of (\ref{mb}) which can then be shown to give
a well-defined expansion for the scalar exchange graph of Fig.2 in the
direct channel. This expansion can
be matched with a conformally invariant OPE
\cite{dhoker,HPR}. However, one finds that in the explicit result for
(\ref{mb}) terms similar to the
second term on the r.h.s. of (\ref{dchann}) are missing
\cite{freedman1,liu,dhoker,HPR}. 
In view of the role played by such terms (shadow
singularities), for the analyticity of the standard CFT$_d$ exchange
graphs it is important to check the analyticity of the AdS$_{d+1}$
graph  in Fig.2 in the crossed channel. We choose for clarity to study
the crossed channel which is obtained from (\ref{mb}) by
the interchange $u\leftrightarrow v$. In this case, the argument of
the $_2F_1$ hypergeometric function becomes $(1-u)$ and we again need to use
degenerate Kummer's relations to analytically continue it into a
series in the variable $u$. Furthermore, the generalized
hypergeometric $_3F_2$ function 
in (\ref{mb}) is {\it Saalschutzian} \cite{slater} and we can use the
non-terminating form of Saalschutz's theorem (Eq. 4.3.4.2 of \cite{slater})
to write it as a sum of a ratio of $\Gamma$-functions and another $_3F_2$
function. Then, after some algebra which involves calculating the Mellin-Barnes
integral we obtain
\ba
\label{adscr}
C(x_1,x_4;x_2,x_3) & = &
\frac{k_1}{(x_{12}^2x_{34}^2)^{\tD}} \Biggl[  
  \sum_{m=0}^{\infty} \frac{u^m}{(m!)^2}{\cal
    D}_m(\frac{\partial}{\partial\xi}) \Bigl( v^{\frac12 \D-\tD}
  f_1(v,\xi) -f_2(v,\xi) -f_3(v,\xi)\Bigl)\Biggl]_{\xi=0}\\
f_1(v,\xi) &=& \frac{\Gamma^2(\frac12 \D+m+\xi)}{\Gamma(
  \D-\frac12 d+1)} 
  {}_2F_1(\sfrac12 \D+m+\xi,\sfrac12 \D+m+\xi; \D-\sfrac12 d+1;v
  ) \label{f1}\,,\\
f_2(v,\xi) & = & \frac{\Gamma^2(\tD+m+\xi)}{\Gamma(\frac12
  \D+\tD-\frac12 d+1)\Gamma(\tD-\frac12 \D+1)}\nonumber \\
 & & \times {}_3F_2\left( \tD+m+\xi,
  \tD+m+\xi, 1;\sfrac12 \D+\tD-\sfrac12 d+1,\tD-\sfrac12
  \D+1;v\right),\label{f2}\\
f_3(v,\xi) &= & \frac{\Gamma(2\tD-\frac12 d)}{\Gamma^2( \tD-\frac12\D)
  \Gamma^2(\tD-\frac12 \D-\frac12 d)} \int_{{\cal C}}\frac{\rmd s}{2\pi{\rm i}}
\frac{\Gamma^2(-s)\Gamma^2(\tD+m+s+\xi)}{ (s+1)(\tD+\frac12 \D-\frac12
  d)}v^s  \nonumber \\ 
& & \hspace{-1cm} \times{}_3F_2\left(\sfrac12 \D-\tD+1,\sfrac12
  \D+\tD-\sfrac12 
  d+1+s,1; s+2,\sfrac12 \D+\tD-\sfrac12 d+1;1\right), \label{f3}
\ea
where again the constant $k_1$ is unimportant here. First we show that
$f_3(v,\xi)$ does 
not contain non-analytic terms as
$v\rightarrow 1$. The argument, whose details can be found in
\cite{HPR}, is as follows. The  $f_{3}(v,\xi)$ term has
the form
\beq
\label{genf3}
f_3(v,\xi) = \int_{{\cal C}}\frac{\rmd s}{2\pi{\rm i}}
\,\Gamma^2(-s)\,g(s,\xi)\,v^s\,,
\eeq
for some function $g(s,\xi)$. The analyticity of $f_3(v,\xi)$  at
$v=0$ allows us 
to write the result of the Mellin-Barnes integration in (\ref{genf3}) as 
\beq
\label{f3epsilon}
f_{3}(v,\xi) =
\frac{\partial}{\partial\e}\left[\sum_{n=0}^{\infty}\frac{v^{n-\e}
    g(n-\e,\xi)}{\Gamma^2(n+1-\e)} \right]_{\e=0}\,.
\eeq
One way to obtain (\ref{f3epsilon}) is to shift by an infinitesimal
parameter $\epsilon$ one of the two $\Gamma$-functions in
(\ref{genf3}) in order to regularize the double poles. The possible
non-analytic terms as $v\rightarrow 1$ in 
(\ref{f3epsilon})  are determined by the large-$n$
asymptotics of the ratio $g(n-\e,\xi)/\Gamma^2(n+1-\e)$. This in turn
can be found using, for example, the Stirling formula
(\ref{stirl}) below and an asymptotic expansion for $_3F_2$ obtained
from Eq. 4.3.4.2 of \cite{slater} by setting (in Slater's notation)
$c=\frac12 \D+\tD-\frac12 d+1+s$, $e=s+2$. In this way we obtain
\beq
\label{goverG}
\left. \frac{g(n-\e,\xi)}{\Gamma^2(n+1-\e)}\right|_{n\rightarrow
  \infty} \approx \sum_{i=1}^2 \sum_{\lambda}^{\infty} 
\tilde{\sigma}_{\lambda ,i}
\frac{\Gamma(A_i+n+1-\e-\lambda)}{ \Gamma(n+1-\e)}\,,
\eeq
for some parameters $A_i$, which depend among others on $m$ and
$\xi$. The coefficients $\tilde{\sigma}_{\lambda,i}$ can in principle be
explicitly determined in terms of the Bernoulli numbers (see below),
but in this case we only require their existence. Then, by
virtue of (\ref{goverG}) we obtain the non-analytic terms
\ba
\label{f3final}
 \left. \sum_{n=0}^{\infty}\frac{v^{n-\e}
    g(n-\e,\xi)}{\Gamma^2(n+1-\e)}\right|_{n.a.} &
\approx & \left. \sum_{i=1}^2\sum_{\lambda}^{\infty} \tilde{\sigma}_{\lambda,i}
\,v^{-\e}\frac{\Gamma(A
  +1-\e-\lambda)}{\Gamma(1-\e)}{}_2F_1(A_i+1-\e-\lambda,1;1-\e;v)
\right|_{n.a.} 
\nonumber \\
& \approx & \sum_{i=0}^2\sum_{\lambda}^{\infty} \tilde{\sigma}_{\lambda,i}
\,\Gamma(A_i+1-\lambda) (1-v)^{-A_i-1+\lambda} \,,
\ea  
where to get the second line of (\ref{f3final}) we used a Kummer
transformation \cite{grad}. The crucial point is now that the 
non-analytic terms are independent of the parameter $\e$ and
therefore  drop out  
when we substitute (\ref{f3final}) into (\ref{f3epsilon}).
 
Next, by yet another Kummer transformation
it is easy to see that the
$f_1(v,\xi)$ term gives the following non-analytic contribution to
(\ref{adscr})
\beq
\label{naf1}
v^{\frac12 \D-\tD}(1-v)^{1-\frac12 d-2m-2\xi}\Gamma(\sfrac12
d-1+2m+2\xi) \sum_{n=0}^{\infty}\frac{
  (1-v)^n}{n!}\frac{(\frac12 \D-\frac12 d+1-m-\xi)_{n}^2}{(2-\frac12
  d-2m-2\xi)_{n}}\,, 
\eeq 
which is similar to the one found in (\ref{cchann}).
It therefore remains
to see whether the $f_2(v,\xi)$ term involves a non-analytic part
which cancels the above contribution. To this end, we would like to
exploit some kind of ``Kummer-like'' transformations for the generalized
hypergeometric function $_3F_2$ in (\ref{f2}).  However, it seems that such
transformations do not exist in general. For this reason we use again
an asymptotic expansion argument as follows. The relevant term in
(\ref{f2}) reads 
\beq
\label{naf2}
\sum_{n=0}^{\infty} \frac{v^n}{n!}\frac{\Gamma^2(\tD+m+\xi+n)
  \Gamma(n+1)}{ \Gamma(\frac12 \D+\tD-\frac12 d+1+n)\Gamma(
  \tD-\frac12 \D +1+n)}\,,
\eeq
and we have extracted an inessential common constant factor from
(\ref{naf1}) and (\ref{naf2}). In order to study the existence of 
non-analytic terms as $v\rightarrow 1$ in (\ref{naf2}) we can use the
identity
\beq
\label{bin}
\sum_{n=1}^{\infty}\frac{\Gamma(a+n+1)}{\Gamma(n+1)}\,v^n =
\Gamma(a+1)\, (1-v)^{-a-1}\,.
\eeq
To do this, we first need to transform (\ref{naf2}) into the form
(\ref{bin}). This can be accomplished with the help of the Stirling
formula \cite{grad} which gives the asymptotics of the ratio
\ba
\label{stirl}
\left.\frac{\Gamma(a+r+1)}{\Gamma(r+1)}\right|_{r\rightarrow \infty}
&\approx& \exp\left[a\ln(r+1)
    +\sum_{k=1}^{\infty}\frac{{\cal P}_{k+1}(a)}{(r+1)^k}\right]\,,
\\
{\cal P}_{k+1}(a) & = & \frac{(-1)^{k+1}}{k}\sum_{l=0}^{a-1}{l^k}\,. 
\label{Pi}
\ea
Using (\ref{stirl}) we can then write
\beq
\label{stappl}
\left. \frac{\Gamma^2(\tD+m+\xi+n)
  }{ \Gamma(\frac12 \D+\tD-\frac12 d+1+n)\Gamma(
  \tD-\frac12 \D +1+n)}\right|_{n\rightarrow \infty} \approx
\sum_{\lambda=0}^{\infty} \sigma_{\lambda}
\frac{\Gamma(\beta-\lambda+n+1)}{ \Gamma(n+1)}\,,
\eeq
where the coefficients $\sigma_{\lambda}$ are recursively obtained
from
\ba
\label{sigmas}
& & \sum_{\lambda=0}^{\infty}\frac{\sigma_{\lambda}}{(n+1)^{\lambda}}
  \exp\left[\sum_{k=1}^{\infty} \frac{{\cal
      P}_{k+1}(\beta-\lambda) }{(n+1)^{k}}\right] =
\exp\left[\sum_{k=1}^{\infty} 
    \frac{2{\cal P}_{k+1}(t_1)-{\cal P}_{k+1}(t_2)
      -{\cal P}_{k+1}(t_3)}{(n+1)^{k}} \right],\\
& & t_1 =  \tD+m+\xi-1\,,\,\,\,t_2= \tD-\sfrac12
\D\,,\,\,\,t_3= \sfrac12 \D +\tD-\sfrac12 d \,,\\
& & \beta  =  2t_1-t_2-t_3 = \sfrac12 d-2+2m+2\xi\,,
\ea 
by matching the powers of $1/(n+1)$ on both sides of
(\ref{sigmas}). Then, using (\ref{stappl}) and (\ref{sigmas}) we are
able to extract the possible non-analytic behavior of (\ref{naf2}),
which is essentially dictated by its  large-$n$ behavior,
to obtain 
\ba 
\label{naf20}
\left. \sum_{n=0}^{\infty} \frac{v^n}{n!}\frac{\Gamma^2(\tD+m+\xi+n)
  \Gamma(n+1)}{ \Gamma(\frac12 \D+\tD-\frac12 d+1+n)\Gamma(
  \tD-\frac12 \D +1+n)}\right|_{n.a.} & \approx &
\sum_{\lambda}^{\infty} \sigma_{\lambda} 
\left[ \sum_{n=0}^{\infty} \frac{\Gamma(\beta
    -\lambda+n+1)}{\Gamma(n+1)} v^n \right]\nonumber \\
 &  & \hspace{-5cm}\approx\sum_{\lambda=0}^{\infty} \sigma_{\lambda}
 \,\Gamma(\sfrac12 d -1 
 +2m+2\xi-\lambda)(1-v)^{1-\frac12 d-2m-2\xi +\lambda}\,.
\ea
Notice that our approach does not determine the analytic part of
(\ref{naf2}). Therefore, from (\ref{adscr}), (\ref{naf1}) and
(\ref{naf20}) the cancellation of non-analytic terms requires that the
equal powers of $1-v$ coincide to all orders in both sides of the
following identity 
\beq
\label{final}
\sum_{\lambda=0}^{\infty} \sigma_{\lambda}
\frac{(-1)^{\lambda}(1-v)^{\lambda}}{ (2-\frac12 d-2m-2\xi)_{\lambda}} =
\sum_{k,l=0}^{\infty} \frac{(1-v)^{k+l}}{k!l!}\frac{(\tD-\frac12
  \D)_{l}(\frac12 \D-\frac12 d+1-m-\xi)_k^2}{ (2-\frac12 d-2m-2\xi)_k}\,.
\eeq
This highly non-trivial identity, which involves on the l.h.s. among
others the Bernoulli numbers through the polynomials (\ref{Pi}),
is analytically proven in \cite{HPR} and it corresponds to
``Kummer-like'' identities for the generalized hypergeometric
$_3F_2$ function. Its first few lower orders can be also shown to be
satisfied 
with the help of e.g. a simple MAPLE algorithm \cite{HPR}.

Concluding, we outlined our proof that the
possible non-analytic terms in AdS$_{d+1}$
scalar exchange graphs drop out by virtue of some highly non-trivial
identities. This is in contrast with the case of scalar exchange
graphs in standard CFT where the cancellation of the non-analytic
terms was due to the appearance of the shadow singularities in the
direct channel. Our result lends support to the idea that the
amplitudes obtained from AdS/CFT correspondence can be analyzed using
conformal OPE techniques \cite{dhoker}. Such an analysis, which also
includes more technical details of our proof here, is presented in
\cite{HPR}.

\section*{Acknowledgments}
The work of A. C. P. is supported by an Alexander von Humboldt
  Fellowship.


\begin{thebibliography}{99}

\bibitem{fradkin}
E. S. Fradkin and M. Ya. Palchik, Phys. Rep. {\bf 44} (1978) 249.

\bibitem{ruehl1}
K. Lang and W. R\"uhl, Nucl. Phys. {\bf B377} (1992) 371.


\bibitem{tassos1}
A. C. Petkou,  ``Conserved currents, consistency relations
and operator product expansions in the conformally invariant $O(N)$
vector model for $2<d<4$'', Ann. Phys. {\bf 249} (1996) 180,
hep-th/9410093; ``Operator product expansions and consistency
relations in a $O(N)$ invariant fermionic CFT for $2<d<4$'',
Phys. Lett. {\bf  B389} (1996) 18, hep-th/9602054. 

\bibitem{depp}
M. D'Eramo, G. Parisi and L. Peliti, Lett. Nuovo Cim. {\bf 2} (1971), 878.

\bibitem{mack}
G. Mack and I. T. Todorov, Phys. Rev. {\bf D8} (1973) 1764;\\
V. K. Dobrev, V. B. Petkova, S. G. Petrova and I. T. Todorov,
Phys. Rev. {\bf D13} (1976) 887.

\bibitem{HPR}
L. Hoffmann, A. C. Petkou and W. R\"uhl, ``{\it Aspects of the conformal
operator product expansion in AdS/CFT correspondence}'', to appear. 

\bibitem{symanzik}
K. Symanzik, Lett. Nuovo Cim. {\bf 3} (1972), 734.

\bibitem{parisi}
S. Ferrara, A. F. Grillo and G. Parisi, Lett. Nuovo Cim. {\bf 5}
(1972) 147; 
S. Ferrara, A. F. Grillo, R. Gatto and G. Parisi, Lett. Nuovo
Cim. {\bf 4} (1972) 115;
S. Ferrara and G. Parisi, Nucl. Phys. {\bf B42} (1972) 281.

\bibitem{koller}
K. Koller, Commun. Math. Phys. {\bf 40} (1975) 15.


\bibitem{grad}
I. S. Gradsteyn and I. M. Ryshik, {\it Table of Integrals, Series and
  Products}, 5th Ed. Academic Press (1994).

\bibitem{klebanov}
I. Klebanov and E. Witten, ``AdS/CFT correspondence and symmetry breaking'',
Nucl. Phys. {\bf B556} (1999) 89, hep-th/9905104.

\bibitem{freedman1}
D.Z.~Freedman, S.D.~Mathur, A.~Matusis and L.~Rastelli,
``Comments on 4-point functions in the CFT/AdS correspondence,"
Phys. Lett. {\bf B452} (1999) 61, hep-th/9808006;
E. D'Hoker and D. Z. Freedman, `General scalar
exchange in $AdS_{d+1}$', Nucl. Phys. {\bf B550} (1999) 261, hep--th/9811257;
E.~D'Hoker, D.Z. Freedman, S.D. Mathur, A. Matusis, L. Rastelli,
``Graviton and gauge boson propagators in $AdS_{d+1}$,"
Nucl. Phys. {\bf B562} (1999) 330,
hep-th/9902042;
E.~D'Hoker, D.Z.~Freedman, S.D.~Mathur, A.~Matusis and L.~Rastelli,
``Graviton exchange and complete 4-point functions in the AdS/CFT
correspondence,'' Nucl. Phys. {\bf B562} (1999) 353,
hep-th/9903196;
E.~D'Hoker, D.Z.~Freedman and L.~Rastelli,
``AdS/CFT 4-point functions: How to succeed at z-integrals without
really trying,'' Nucl. Phys. {\bf B562} (1999) 395,
hep-th/9905049.

\bibitem{liu}
H. Liu, `Scattering in Anti--de Sitter space and operator
product expansion', Phys. Rev. {\bf D60} 106005 (1999), hep--th/9811152.

\bibitem{maldacena}
J. Maldacena, ``The large N limit of superconformal field 
theories and supergravity'', Adv. Theor. Math. Phys. {\bf 2} (1998) 
231, hep-th/9711200; 
S.S. Gubser, I.R. Klebanov and A.M. 
Polyakov, ``Gauge theory correlators from noncritical string 
theory'', Phys. Lett. {B428} (1998) 105, hep-th/9802109; 
E. Witten, 
``Anti-de  Sitter space and holography'', Adv. Theor. Math. 
Phys. {\bf 2} (1998) 253, hep-th/9802150;
O.~Aharony, S.S.~Gubser, J.~Maldacena, H.~Ooguri and Y.~Oz,
``Large N field theories, string theory and gravity'',
hep-th/9905111.

\bibitem{titsch}
E. C. Titchmarsh, {\it The theory of Functions}, 2nd Ed. Oxford
University Press (1939). (Reprinted 1975).

\bibitem{slater}
L. J. Slater, {\it Generalized Hypergeometric Functions}, Cambridge
University Press (1966).

\bibitem{dhoker}
E.~D'Hoker, S.D.~Mathur, A.~Matusis and L.~Rastelli, ``The operator
product expansion of ${\cal N}$=4 SYM and the 4-point functions of
supergravity'', hep-th/9911222.


\end{thebibliography}
\end{document}